# Coordinated AC/DC Microgrid Optimal Scheduling

Abdulaziz Alanazi, Hossein Lotfi, Amin Khodaei
Department of Electrical and Computer Engineering
University of Denver
Denver, CO 80210, USA
alanazi.alanazi@du.edu, hossein.lotfi@du.edu, amin.khodaei@du.edu

***Abstract*— This paper proposes a coordinated optimal scheduling model for hybrid AC/DC microgrids. The objective of the proposed model is to minimize the total microgrid operation cost when considering interactions between AC and DC sub-systems of the microgrid network. Nonlinear power flow equations for AC and DC networks have been linearized through a proposed model to enable formulating the problem by mixed integer linear programming (MILP) which expedites the solution process and ensures better solutions in terms of optimality. The proposed model is tested on the modified IEEE 33-bus test system. Numerical simulations exhibit the merits of the proposed coordinated AC/DC optimal scheduling model and further analyze its sensitivity to various decisive operational parameters.**

## NOMENCLATURE

***Indices:***

| | |
|---|---|
| $i$ | Index for DERs |
| $m, n$ | Index for buses |
| $t$ | Index for time |

***Sets:***

| | |
|---|---|
| B | Set of buses |
| $B_{ac}$ | Set of AC buses |
| $B_m$ | Set of buses adjacent to bus m |
| C | Set of converters |
| G | Set of dispatchable DGs |
| $G_m$ | Set of dispatchable DGs connected to bus m |
| L | Set of lines |
| $L_{ac}$ | Set of AC lines |
| S | Set of DES units |
| W | Set of nondispatchable DGs (wind and solar) |
| $W_m$ | Set of nondispatchable DGs connected to bus m |

***Parameters:***

| | |
|---|---|
| $b$ | Line susceptance |
| $D$ | DES depth of discharge |
| $DT$ | Minimum down time |
| $E^{max}$ | Maximum energy capacity of DES |
| $g$ | Line conductance |
| $P^{max}$ | Maximum generation limit |
| $P^{min}$ | Minimum generation limit |
| $P^{M,max}$ | Line capacity between utility and microgrid |
| $PC^{max}$ | Maximum real power for converters |
| $PD$ | Load active power |
| $PL^{max}$ | Maximum active power flow for distribution lines |
| $QC^{max}$ | Maximum reactive power for converters |
| $QD$ | Load reactive power |
| $QL^{max}$ | Maximum reactive power flow for distribution lines |
| $RD$ | Ramp down rate |
| $RU$ | Ramp up rate |
| $UT$ | Minimum up time |
| $w$ | Line type (0 when AC, 1 when DC) |
| $\rho$ | Electricity market price |
| $\eta$ | DES efficiency |

***Variables:***

| | |
|---|---|
| $E$ | DES stored energy |
| $F(.)$ | Generation cost |
| $I$ | Commitment state of dispatchable DGs |
| $P$ | Active power of DGs |
| $P^B$ | DES output power |
| $P^{ch}$ | DES charging power |
| $P^{dch}$ | DES discharging power |
| $P^M$ | Exchanged power with the utility grid |
| $PC$ | Active power of converters |
| $PL$ | Active power of lines |
| $Q$ | Reactive power of DGs |
| $QC$ | Reactive power of converters |
| $QL$ | Reactive power of lines |
| $T^{OFF}$ | Number of successive OFF hours |
| $T^{ON}$ | Number of successive ON hours |
| $u$ | DES charging state |
| $v$ | DES discharging state |
| $V$ | Bus voltage magnitude |
| $y$ | Startup indicator of dispatchable DGs |
| $z$ | Shutdown indicator of dispatchable DGs |
| $\theta$ | Bus voltage angle |

## I. INTRODUCTION

The variety applications of microgrids along with the many promising benefits, such as energy efficiency, improved reliability and power quality, and facilitated integration of renewable resources, have resulted in considering this technology as a main constituent of future power systems [1-5]. Most microgrids are designed to be AC, similar to the traditional approach to develop and expand power systems. However, after the widespread use of DC loads, such as data and communication centers, electronic devices, and electric vehicles, on one hand, and deployment of DC distributed energy resources (DERs), such as solar PV, fuel cell, and distributed energy storage (DES), on the other hand, DC microgrids would potentially become more economical to deploy. There are many advantages associated with DC microgrids such as no need for synchronization of DERs, easier integration of DC components, and higher system efficiency due to elimination of multiple AC-DC converters. Hybrid AC/DC microgrids are more desirable when both AC and DC loads/DERs exist in the system. Hybrid microgrids would take advantage of both AC and DC microgrids by deployment of AC and DC DERs/loads. In these microgrids, there are two main bus types, i.e., AC and DC, in which AC components are connected to AC buses and DC components to DC buses. There is a need for bidirectional converters to connect AC and DC buses in the microgrid [6,7]. According to the level of generation and loads in different buses in AC and DC networks, power can go from AC to DC network (where the converter acts as a rectifier) or from DC to AC (where the converter acts as an inverter). The microgrid can be connected to the utility grid from both AC and DC sides. One impressive feature of microgrid that increases its reliability and resilience is the islanding capability which enables it to be disconnected from the utility grid in case of faults or disturbance in the upstream utility grid [7,8]. The energy management in microgrids is very important both economically and energy wise. It is more complex when there is a hybrid AC/DC microgrid instead of individual AC or DC microgrid.

The study in [9] proposes an optimal day-ahead scheduling for microgrid participation in frequency regulation markets. The balance between energy and ancillary services for all distributed generations (DGs) in the microgrid is determined in order to participate in frequency regulation markets. A bidirectional power flow control in a DC microgrid is presented in [10]. A bidirectional hybrid DC-DC converter links two DC voltage buses in this model. A current controller is used beside the converter designed in frequency domain working based on the Bode plot. The study in [11] presents an optimal power flow for power management in microgrids. The optimal location of microgrid in a distribution network is determined by considering different penetration ratios of microgrid. The optimal microgrid location for two test distribution systems is determined in this study, and voltage profile and line losses are compared to each other. In [12], a new control algorithm for power flow management in combined AC/DC microgrids is proposed. There is an interlinking converter to connect AC and DC parts of the microgrid which is controlled using the method presented in this study. It is shown that droop based control algorithm reduces multiple power conversion stages. The study in [13] presents a multi-purpose interlinking converter control for multiple hybrid AC/DC microgrid operations. The objective of the proposed model is to optimally share the active/reactive power among multiple microgrids. Interlinking converter in each microgrid is equipped with PD compensated droop controller. The simulation results on two islanded hybrid AC/DC microgrids show that the proposed model provides a robust performance under various scenarios. In [14], a microgrid optimal scheduling model is proposed considering multi-period islanding constraints. The minimization of microgrid total operation cost is taken as the objective, and Benders decomposition method has been used to expedite the running process. The scheduling problem is decomposed into a master problem (grid-connected operation) and a subproblem (islanded operation). These two problems are coordinated using islanding cuts. Numerical simulations show that islanding criterion would improve reliability whereas increases the operation cost. A resiliency-oriented microgrid optimal scheduling model is presented in [15] in which the objective is to minimize the microgrid load curtailment by optimal scheduling of DERs in the event of power supply interruption from the utility grid. The study in [4] presents a decentralized operation framework for hybrid AC/DC microgrids. There is a bidirectional AC/DC converter to link AC and DC networks of the hybrid microgrid. Decomposition method has been used to decompose the AC and DC scheduling problems in which first the AC optimal scheduling problem is solved. If the solution converges, the DC scheduling problem is solved.

In this paper, an optimal scheduling model for coordinated AC/DC microgrids is presented. The power flow equations for AC and DC networks are linearized using some approximations in order that the problem could be formulated by mixed integer linear programming (MILP). The scheduling problem for the whole microgrid, including AC and DC sub-systems, is formulated and solved in an integrated fashion.

The rest of the paper is organized as follows. Section II presents the coordinated AC/DC microgrid optimal scheduling model outline and formulation. Numerical results are represented in Section III. Section IV provides discussions, and Section V concludes the paper.

## II. Model Outline and Formulation

The proposed model aims at developing an optimal scheduling framework for coordinated AC/DC microgrid. This objective is achieved by solving the power flow for both AC and DC networks together. AC network is linked to the DC network by bidirectional converters, so the power converted from AC to DC or vice versa is also considered in the proposed model. The converters' efficiency and the power flow direction are also modeled in formulating the problem. Moreover, the power losses in the distribution lines are considered. Since the model is proposed for microgrid operation, the problem is solved for a 24-hour period. The objective function is proposed in (1) which minimizes the total operation cost. The first term in (1) represents the operation cost of dispatchable DGs and the second term denotes the cost or benefit of exchanged power with the utility grid. The exchanged power might be positive (i.e., microgrid purchases power from the utility grid when the imported power is less expensive than local generation), or negative (i.e., microgrid sells power to the utility grid when it offers a higher price compared to the local generation).

$$\min\left(\sum_{i\in G}\sum_{t}F_i(P_{it})+\sum_{t}\rho_t P_t^M\right) \tag{1}$$

The objective function is subject to a number of linear power flow constraints (2)-(19), dispatchable DGs' operation constraints (20)-(25), and DES constraints (26)-(31).

### A. Linear Power Flow Constraints

The nonlinear power flow equations between buses *m* and *n* in an AC system are shown in (2) and (3). By assuming (4)-(7), the nonlinear power flow equations can be linearized:

$$PL_{mnt} = g_{mn}V_{mt}^2 - V_{mt}V_{nt}(g_{mn}\cos(\theta_{mt}-\theta_{nt})) - V_{mt}V_{nt}(b_{mn}\sin(\theta_{mt}-\theta_{nt})) \quad \forall mn \in \mathrm{L}, \forall t \tag{2}$$

$$QL_{mnt} = -b_{mn}V_{mt}^2 - V_{mt}V_{nt}(b_{mn}\cos(\theta_{mt}-\theta_{nt})) - V_{mt}V_{nt}(g_{mn}\sin(\theta_{mt}-\theta_{nt})) \quad \forall mn \in \mathrm{L}, \forall t \tag{3}$$

$$\sin(\theta_{mt}-\theta_{nt}) \approx \theta_{mt}-\theta_{nt} \quad \forall mn \in \mathrm{L_{ac}}, \forall t \tag{4}$$

$$\cos(\theta_{mt}-\theta_{nt}) \approx 1 \quad \forall mn \in \mathrm{L_{ac}}, \forall t \tag{5}$$

$$V_{mt} = 1.0 + \Delta V_{mt} \quad \forall m, \forall t \tag{6}$$

$$\theta_{mt} = 0 + \Delta\theta_{mt} \quad \forall m \in \mathrm{B_{ac}}, \forall t \tag{7}$$

Linear power flow equations for active and reactive power are obtained and respectively represented by (8) and (9):

$$PL_{mnt} = g_{mn}(\Delta V_{mt}-\Delta V_{nt}) - b_{mn}(\Delta\theta_{mt}-\Delta\theta_{nt}) + g_{mn}\Delta V_{mt}(\Delta V_{mt}-\Delta V_{nt}) \quad \forall mn \in \mathrm{L_{ac}}, \forall t \tag{8}$$

$$QL_{mnt} = -b_{mn}(\Delta V_{mt}-\Delta V_{nt}) - g_{mn}(\Delta\theta_{mt}-\Delta\theta_{nt}) - b_{mn}\Delta V_{mt}(\Delta V_{mt}-\Delta V_{nt}) \quad \forall mn \in \mathrm{L_{ac}}, \forall t \tag{9}$$

These equations are further modified to include the DC power flow considering the line type $w_{mn}$ (0 when line is AC and 1 when it is DC), as shown in (10) and (11). It should be noted that the third term in (10) represents power losses in distribution lines. Also, it is worth mentioning that reactive power flow is zero for the DC sub-system since DC network has no reactive power.

$$PL_{mnt} = g_{mn}(\Delta V_{mt}-\Delta V_{nt}) - b_{mn}(\Delta\theta_{mt}-\Delta\theta_{nt})(1-w_{mn}) + g_{mn}\Delta V_{mt}(\Delta V_{mt}-\Delta V_{nt}) \quad \forall mn \in \mathrm{L}, \forall t \tag{10}$$

$$QL_{mnt} = \begin{pmatrix} -b_{mn}(\Delta V_{mt}-\Delta V_{nt}) - g_{mn}(\Delta\theta_{mt}-\Delta\theta_{nt}) \\ -b_{mn}\Delta V_{mt}(\Delta V_{mt}-\Delta V_{nt}) \end{pmatrix}(1-w_{mn}) \quad \forall mn \in \mathrm{L}, \forall t \tag{11}$$

Line real and reactive power flow limits are represented by (12) and (13), respectively. Equation (14) represents the limit of voltage magnitude for each bus in the microgrid. The converters' real and reactive power cannot exceed their limits (15), (16). The nodal load balance equation for active powers (17) ensures that the total real power from DGs (dispatchable and nondispatchable), DES, lines, and converters, plus the exchanged real power with utility grid equal the real power load demand at each bus. On the other hand, the nodal load balance equation for reactive power (18) guarantees the total reactive power from DGs, lines, converters, and the exchanged reactive power with utility grid equals the reactive power load demand at each bus only in the AC side of the microgrid. The exchanged power with the utility grid cannot exceed the capacity of the line connecting the utility grid to the microgrid (19).

$$-PL_{mn}^{\max} \le PL_{mnt} \le PL_{mn}^{\max} \quad \forall mn \in \mathrm{L}, \forall t \tag{12}$$

$$-QL_{mn}^{\max} \le QL_{mnt} \le QL_{mn}^{\max} \quad \forall mn \in \mathrm{L}, \forall t \tag{13}$$

$$V_m^{\min} \le V_{mt} \le V_m^{\max} \quad \forall m \tag{14}$$

$$-PC_{mn}^{\max} \le PC_{mnt} \le PC_{mn}^{\max} \quad \forall mn \in \mathrm{C}, \forall t \tag{15}$$

$$-QC_{mn}^{\max} \le QC_{mnt} \le QC_{mn}^{\max} \quad \forall mn \in \mathrm{C}, \forall t \tag{16}$$

$$\sum_{i\in\{\mathrm{G,W}\}} P_{it} + \sum_{i\in \mathrm{S}} P_{it}^B + \sum_{n\in \mathrm{B_m}} PL_{mnt} + \sum_{n\in \mathrm{C}} PC_{mnt} + P_t^M = PD_{mt} \quad \forall m, \forall t \tag{17}$$

$$\sum_{i\in\{\mathrm{G_m,W_{m,ac}}\}} Q_{it} + \sum_{n\in \mathrm{B_m}} QL_{mnt} + \sum_{n\in \mathrm{C}} QC_{mnt} + Q_t^M = QD_{mt} \quad \forall m \in \mathrm{B_{ac}}, \forall t \tag{18}$$

$$-P^{M,\max} \le P_t^M \le P^{M,\max} \quad \forall t \tag{19}$$

### B. DG Constraints

The dispatchable DGs' operation constraints are modeled in (20)-(25). The capacity limits of real and reactive powers for dispatchable DGs are respectively modeled by (20) and (21) considering the commitment state *I*, as it will be one when the unit is ON, otherwise it is zero. The ramping up and down rate limits are represented by (22)-(23). The minimum up and down times for each DG are represented by (24)-(25). The startup state *y* is one only when the unit is started up, otherwise it is zero. Similarly, the shut down state *z* will be one only when the unit is shut down, otherwise it is zero.

$$P_i^{\min} I_{it} \le P_{it} \le P_i^{\max} I_{it} \quad \forall i \in \mathrm{G}, \forall t \tag{20}$$

$$Q_i^{\min} I_{it} \le Q_{it} \le Q_i^{\max} I_{it} \quad \forall i \in \mathrm{G}, \forall t \tag{21}$$

$$P_{it} - P_{i(t-1)} \le RU_i \quad \forall i \in \mathrm{G}, \forall t \tag{22}$$

$$P_{i(t-1)} - P_{it} \le RD_i \quad \forall i \in \mathrm{G}, \forall t \tag{23}$$

$$T_{it}^{ON} \ge UT_i z_{i(t+1)} \quad \forall i \in \mathrm{G}, \forall t \tag{24}$$

$$T_{it}^{OFF} \ge DT_i y_{i(t+1)} \quad \forall i \in \mathrm{G}, \forall t \tag{25}$$

### C. DES Constarints

The discharging power (26) is always positive since the DES is producing power while it is discharging. Conversely, the charging power (27) is negative as the DES is consuming power when it is charging. The DES output power is the summation of charging and discharging powers (28). The hourly stored energy is calculated in (29) as the stored energy at previous hour minus the hourly charged or discharged power, so the stored energy will increase when DES is charging (as the charging power is negative) and will decrease when DES is discharging (as the discharging power is positive). The stored energy is limited by (30) considering DES depth of discharge. The binary discharging state *v* is set to 1 when the DES is discharging, otherwise it is set to 0. Similarly, the binary charging state *u* is set to 1 when the DES is charging, otherwise it is set to 0. By using (31), it is ensured that both binary variables are not equal to 1 at the same time as the DES does not charge and discharge simultaneously.

$$0 \le P_{it}^{dch} \le P_i^{\max} v_{it} \quad \forall i \in \mathrm{S}, \forall t \tag{26}$$

$$-P_i^{\max} u_{it} \le P_{it}^{ch} \le 0 \qquad \forall i \in S, \forall t \qquad (27)$$

$$P_{it}^{B} = P_{it}^{dch} + P_{it}^{ch} \qquad \forall i \in S, \forall t \qquad (28)$$

$$E_{it} = E_{i(t-1)} - \frac{P_{it}^{dch}}{\eta_i} - P_{it}^{ch} \qquad \forall i \in S, \forall t \qquad (29)$$

$$(1-D)E_i^{\max} \le E_{it} \le E_i^{\max} \qquad \forall i \in S, \forall t \qquad (30)$$

$$u_{it} + v_{it} \le 1 \qquad \forall i \in S, \forall t \qquad (31)$$

## III. Numerical Simulations

The proposed model is applied to a modified IEEE 33-bus test system, as shown in Fig. 1. This system has 33 buses, 30 distribution lines, and 32 loads. To consider it as a microgrid, 3 thermal generation units, 1 solar PV, and 1 DES are added to this network, with adequate capacity to enable an islanded operation. Tables I, II, and III show the characteristics of DGs, DES and distribution lines, respectively. The lines between buses 3-23 and 6-26 are removed and replaced by bidirectional converters to have AC and DC sub-systems. Moreover, the DC sub-system is connected to the utility grid at bus 23, whereas AC sub-system is connected to the utility grid at bus 1. Efficiencies of rectifiers and inverters in the test system are considered to be 97% and 93%, respectively. The problem is formulated by MIP and solved using CPLEX 12.6 [16].

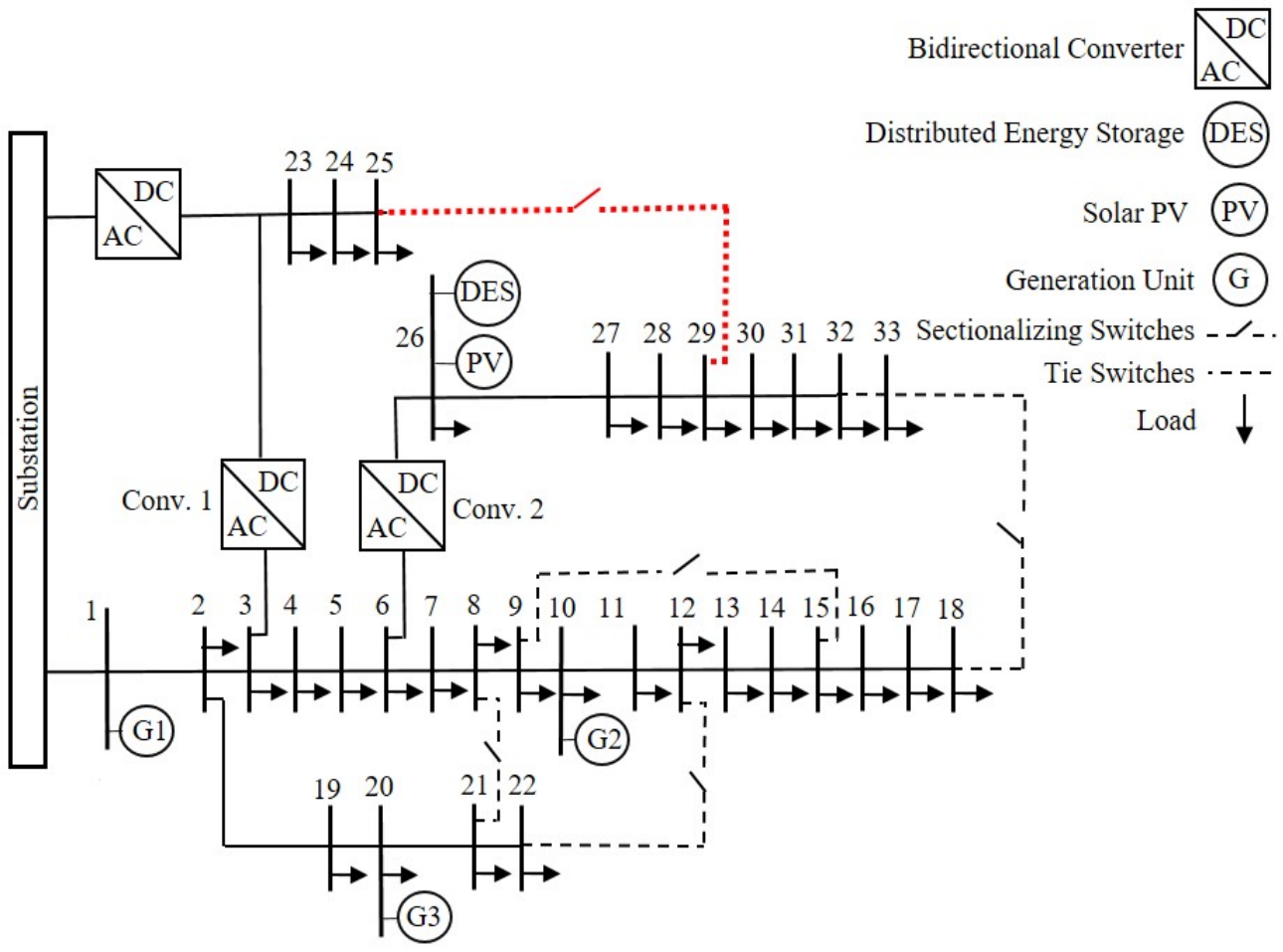

Fig. 1. Modified IEEE 33-bus test system.

TABLE I
DGs' CHARACTERISTICS

| Unit | Type | Cost Coefficient ($/kWh) | Min-Max Limits (kW) | Min. Up/Down Time (h) | Ramp Up/Down Rate (kW/h) |
|---|---|---|---|---|---|
| G1 | Thermal | 0.030 | 1000-3000 | 4 | 500 |
| G2 | Thermal | 0.040 | 500-2000 | 5 | 300 |
| G3 | Thermal | 0.060 | 500-1000 | 6 | 300 |
| G4 | PV | 0 | 0-2000 | - | - |

TABLE II
DES CHARACTERISTICS

| Power Rating (kW) | Energy Rating (kWh) | Depth of Discharge (%) | Efficiency (%) |
|---|---|---|---|
| 2000 | 6000 | 80 | 90 |

TABLE III
LINES' CHARACTERISTICS

| Line | Type | From bus | To bus | R(Ω) | X(Ω) | Line Capacity (kW)/(kVAR) |
|---|---|---|---|---|---|---|
| 1 | AC | 1 | 2 | 0.0922 | 0.0470 | 4600 |
| 2 | AC | 2 | 3 | 0.4930 | 0.2511 | 4100 |
| 3 | AC | 3 | 4 | 0.3660 | 0.1864 | 2500 |
| 4 | AC | 4 | 5 | 0.3811 | 0.1941 | 2400 |
| 5 | AC | 5 | 6 | 0.8190 | 0.7070 | 2300 |
| 6 | AC | 6 | 7 | 0.1872 | 0.6188 | 1050 |
| 7 | AC | 7 | 8 | 0.7114 | 0.2351 | 1050 |
| 8 | AC | 8 | 9 | 1.0300 | 0.7400 | 1050 |
| 9 | AC | 9 | 10 | 1.0440 | 0.7400 | 1050 |
| 10 | AC | 10 | 11 | 0.1966 | 0.0650 | 1050 |
| 11 | AC | 11 | 12 | 0.3744 | 0.1298 | 1050 |
| 12 | AC | 12 | 13 | 1.4680 | 1.1550 | 500 |
| 13 | AC | 13 | 14 | 0.5416 | 0.7129 | 450 |
| 14 | AC | 14 | 15 | 0.5910 | 0.5260 | 300 |
| 15 | AC | 15 | 16 | 0.7463 | 0.5450 | 250 |
| 16 | AC | 16 | 17 | 1.2890 | 1.7210 | 250 |
| 17 | AC | 17 | 18 | 0.7320 | 0.5740 | 100 |
| 18 | AC | 2 | 19 | 0.1640 | 0.1565 | 500 |
| 19 | AC | 19 | 20 | 1.5042 | 1.3554 | 500 |
| 20 | AC | 20 | 21 | 0.4095 | 0.4784 | 210 |
| 21 | AC | 21 | 22 | 0.7089 | 0.9373 | 110 |
| 22 | DC | 23 | 24 | 0.8980 | 0.7091 | 1500 |
| 23 | DC | 24 | 25 | 0.8960 | 0.7011 | 1500 |
| 24 | DC | 26 | 27 | 0.2842 | 0.1447 | 1500 |
| 25 | DC | 27 | 28 | 1.0590 | 0.9337 | 1500 |
| 26 | DC | 28 | 29 | 0.8042 | 0.7006 | 1500 |
| 27 | DC | 29 | 30 | 0.5075 | 0.2585 | 1500 |
| 28 | DC | 30 | 31 | 0.9744 | 0.9630 | 500 |
| 29 | DC | 31 | 32 | 0.3105 | 0.3619 | 500 |
| 30 | DC | 32 | 33 | 0.3410 | 0.5302 | 100 |
| 31 | Tie line | 25 | 29 | 0.5000 | 0.5000 | 1500 |

The proposed model is solved for a 24-hour time horizon to obtain the optimal scheduling of the coordinated AC/DC microgrid for the following cases:

**Case 1**: Disconnected AC and DC sub-systems
**Case 2**: Interconnected AC and DC sub-systems by using bidirectional converters
**Case 3:** Impact of putting AC loads in DC sub-system and DC loads in AC sub-system
**Case 4:** Impact of installing AC DERs in DC sub-system and DC DERs in AC sub-system

**Case 1**: In this case, the optimal scheduling problem is solved for disconnected AC and DC sub-systems, as shown in Fig. 2. The dotted line in Fig. 2 is a tie line and is closed in order to connect the two isolated DC feeders. All the loads in AC (DC) sub-system are AC (DC) loads. The total power losses in this case are calculated as 1846.756 kW. Unit G1 is committed at all hours while units G2 and G3 are OFF. The total operation cost is $3250.526. Optimal DER schedule is represented in Table IV. Note that the charging, discharging, and idle states in DES are denoted by -1, 1, and 0, respectively.

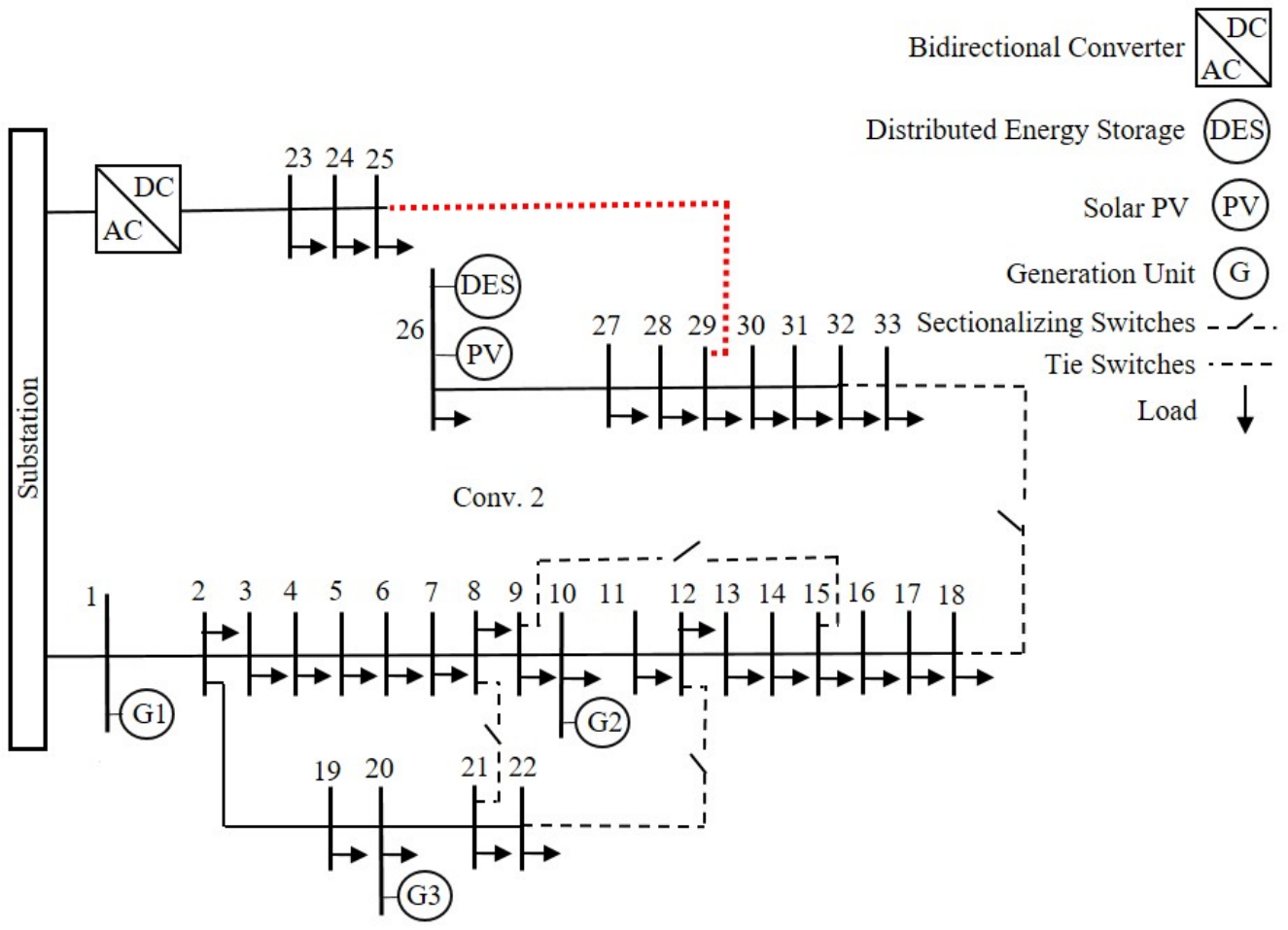

Fig. 2. Disconnected AC and DC sub-systems.

TABLE IV
DERS' OPTIMAL SCHEDULING IN CASE 1

| | Hours (1-24) | | | | | | | | | | | | | | | | | | | | | | | |
|---|---|---|---|---|---|---|---|---|---|---|---|---|---|---|---|---|---|---|---|---|---|---|---|---|
| G1 | 1 | 1 | 1 | 1 | 1 | 1 | 1 | 1 | 1 | 1 | 1 | 1 | 1 | 1 | 1 | 1 | 1 | 1 | 1 | 1 | 1 | 1 | 1 | 1 |
| G2 | 0 | 0 | 0 | 0 | 0 | 0 | 0 | 0 | 0 | 0 | 0 | 0 | 0 | 0 | 0 | 0 | 0 | 0 | 0 | 0 | 0 | 0 | 0 | 0 |
| G3 | 0 | 0 | 0 | 0 | 0 | 0 | 0 | 0 | 0 | 0 | 0 | 0 | 0 | 0 | 0 | 0 | 0 | 0 | 0 | 0 | 0 | 0 | 0 | 0 |
| DES | -1 | -1 | -1 | -1 | -1 | -1 | -1 | -1 | -1 | -1 | -1 | 1 | -1 | -1 | -1 | -1 | -1 | 1 | 1 | 1 | 1 | 1 | 1 | -1 |

**Case 2**: In this case, the optimal scheduling problem is solved for the integrated AC and DC sub-systems. AC and DC networks are connected via converters as shown in Fig. 1. In order to compare this case with the previous one, the AC and DC load ratios are kept the same. The total power loss in this case increases to 2061.34 kW, which represents an increase of 11.62%. However, the total microgrid operation cost decreases by 36.42% to reach $2066.796. Optimal DER schedule is shown in Table V. It can be seen that DES schedule is changed in hours that are highlighted in gray. This result advocates that the interconnected case is more economical than the previous case. Table VI summarizes and compares the results in Cases 1 and 2.

TABLE V
DERS' OPTIMAL SCHEDULING IN CASE 2

| | Hours (1-24) | | | | | | | | | | | | | | | | | | | | | | | |
|---|---|---|---|---|---|---|---|---|---|---|---|---|---|---|---|---|---|---|---|---|---|---|---|---|
| G1 | 1 | 1 | 1 | 1 | 1 | 1 | 1 | 1 | 1 | 1 | 1 | 1 | 1 | 1 | 1 | 1 | 1 | 1 | 1 | 1 | 1 | 1 | 1 | 1 |
| G2 | 0 | 0 | 0 | 0 | 0 | 0 | 0 | 0 | 0 | 0 | 0 | 0 | 0 | 0 | 0 | 0 | 0 | 0 | 0 | 0 | 0 | 0 | 0 | 0 |
| G3 | 0 | 0 | 0 | 0 | 0 | 0 | 0 | 0 | 0 | 0 | 0 | 0 | 0 | 0 | 0 | 0 | 0 | 0 | 0 | 0 | 0 | 0 | 0 | 0 |
| DES | -1 | -1 | -1 | -1 | -1 | 1 | -1 | 1 | -1 | 0 | 0 | 0 | 0 | 0 | 0 | 0 | 1 | 1 | 0 | -1 | 1 | 0 | 0 | 0 |

TABLE VI
SUMMARY OF THE RESULTS IN CASES 1 AND 2

| | Total Power Loss (kW) | Operation Cost ($) |
|---|---|---|
| Case 1 | 1846.756 | 3250.526 |
| Case 2 | 2061.34 | 2066.796 |
| Change | 11.62% increase | 36.42% decrease |

**Case 3:** In this case, the effect of changing the ratio of AC and DC loads in AC and DC sub-systems in the interconnected mode is studied as follows:

3.1. Increasing AC loads in DC sub-system from 0% to 100% by steps of 10%.
3.2. Increasing DC loads in AC sub-system from 0% to 100% by steps of 10%.

The loads in opposite-type buses are supplied via rectifiers or inverters. The operation cost and total power losses are calculated and summarized in Tables VII and VIII. As it can be seen, increasing AC loads in DC buses and DC loads in AC buses would increase the total power losses and operation cost. The increase is monotonic and almost linear, but when the loads in AC sub-system are 100% DC, unit G2 is turned ON. Consequently, the total operation cost sharply increases and the total power losses decrease. As a result, it would be more economical and energy-efficient to locate AC and DC loads in the same-type bus.

TABLE VII
IMPACT OF CHANGING THE RATIO OF AC LOAD IN DC SUB-SYSTEM

| AC Load Ratio (%) | DC Load Ratio (%) | Total Power losses (kW) | Operation Cost ($) |
|---|---|---|---|
| 0 | 100 | 2061.340 | 2066.796 |
| 10 | 90 | 2117.862 | 2149.770 |
| 20 | 80 | 2187.090 | 2234.108 |
| 30 | 70 | 2241.056 | 2317.377 |
| 40 | 60 | 2311.146 | 2401.457 |
| 50 | 50 | 2359.365 | 2485.359 |
| 60 | 40 | 2416.751 | 2569.054 |
| 70 | 30 | 2499.163 | 2654.367 |
| 80 | 20 | 2560.119 | 2738.793 |
| 90 | 10 | 2621.083 | 2824.078 |
| 100 | 0 | 2689.165 | 2909.388 |

TABLE VIII
IMPACT OF CHANGING THE RATIO OF DC LOAD IN AC SUB-SYSTEM

| DC Load Ratio (%) | AC Load Ratio (%) | Total Power losses (kW) | Operation Cost ($) |
|---|---|---|---|
| 0 | 100 | 2061.34 | 2066.796 |
| 10 | 90 | 2092.211 | 2130.141 |
| 20 | 80 | 2128.647 | 2194.063 |
| 30 | 70 | 2150.294 | 2256.571 |
| 40 | 60 | 2184.33 | 2320.816 |
| 50 | 50 | 2210.135 | 2384.562 |
| 60 | 40 | 2246.583 | 2448.671 |
| 70 | 30 | 2267.952 | 2511.845 |
| 80 | 20 | 2306.163 | 2576.528 |
| 90 | 10 | 2336.817 | 2640.701 |
| 100 | 0 | 933.118 | 3850.576 |

**Case 4:** In this case, the effect of installing DC DERs in AC sub-system and AC DERs in DC sub-system in the interconnected mode is investigated as follows:

4.1. Moving DC DERs (i.e., solar PV and DES) to AC sub-system
4.2. Moving AC DERs to DC sub-system
4.3. Moving AC DERs to DC sub-system and DC DERs to AC sub-system simultaneously

Similar to the loads, the DERs in opposite-type buses are connected via rectifiers or inverters. When DC DERs are installed in AC sub-system, the total power losses and operation cost respectively increase by 3.31% and 19.52% compared to Case 2. Whereas, when AC DERs are installed in DC sub-system, the results show a reduction of 29.61% in total power losses and an increase of 176.22% in the operation cost compared to Case 2. Finally, when all DERs are installed in opposite-type buses, the operation cost significantly increases, by 193.28%, while total power losses decrease by 39.48% compared to Case 2. A summary and comparison of the results in Case 4 is illustrated in Table IX. It can be concluded that installing DERs in the same-type network is more economical and efficient than installing them in an opposite-type bus.

TABLE IX
SUMMARY OF CASE 4 RESULTS

| | Total Power Losses (kW) | Total Operation Cost ($) |
|---|---|---|
| DC DERs in AC sub-system | 2129.626 | 2470.200 |
| AC DERs in DC sub-system | 1451.030 | 5708.959 |
| AC DERs in DC sub-system and DC DERs in AC sub-system | 1247.508 | 6061.429 |

## IV. DISCUSSIONS

Key factors obtained from studies on the proposed optimal scheduling framework for coordinated AC/DC microgrids are listed as follows:

- A coordinated scheduling of the AC and DC sub-systems would result in a more economical solution when compared to the separated model, however, coordinated scheduling may potentially result in a higher total power loss.
- Installing loads in opposite-type buses causes an increase in required energy to account for losses in converters, modeled through converters efficiency in the proposed problem. Consequently, the total operation cost would increase. Similarly, installing DERs in opposite-type buses would increase the total operation cost.
- Increasing AC (DC) loads in DC (AC) sub-system may result in commitment of additional units in the microgrid to supply the local loads, especially when there is a congestion in the distribution network and cheaper units exist, or the utility grid cannot fully supply all the loads. This reduces the total power losses and increases the microgrid operation cost.

## V. CONCLUSION

A coordinated optimal scheduling model for hybrid AC/DC microgrids was proposed in this paper. The nonlinear power flow equations were linearized and integrated to the optimal scheduling model to be further formulated as MILP. The proposed model was tested on the modified IEEE 33-bus test system with three thermal generation units, one solar PV and one DES. Four cases were studied, where in Cases 1 and 2, the optimal scheduling problem for disconnected and interconnected sub-systems was investigated. The comparison between the two cases showed that using the proposed coordinated model would be more economical as the operation cost considerably decreased. The coordinated model, however, was not necessarily more energy-efficient, as the losses in this case increased. It was further shown through Cases 3 and 4 that placing AC and DC resources (DERs and loads) in the opposite-type network would increase energy losses and operation costs, meaning that locating resources in the same-type network would be more energy and cost efficient.